\documentclass[referee, useAMS, usenatbib, amstex, amssymb]{mn2e}
\usepackage{psfig}
\usepackage{graphicx}
\usepackage{epsf}
\usepackage{bm}
\usepackage{lscape}

\title [Metallicity Gradient from RAVE DR3]
{Local stellar kinematics from RAVE data:
III. Radial and Vertical Metallicity Gradients based on Red Clump Stars}

\author[Bilir et al.]
       {S. Bilir $^{1}$\thanks{E-mail: sbilir@istanbul.edu.tr},
S. Karaali$^{1}$\thanks{Retired}, S. Ak$^{1}$, \"O. \"Onal$^{1}$, N. D. Da\u gtekin$^{1}$, T. Yontan$^{1}$,
\newauthor
 G. Gilmore$^{2,3}$, G. M. Seabroke$^{4}$
\\
  $^1$Istanbul University, Science Faculty, Department of Astronomy and Space
Sciences, 34119, University-Istanbul, Turkey\\
  $^2$Institute of Astronomy, Madingley Road, Cambridge, CB3 OHA, UK\\
  $^3$Astronomy Department, Faculty of Science, King Abdulaziz University, P.O.
Box 80203, Jeddah 21589, Saudi Arabia\\
  $^4$Mullard Space Science Laboratory, University College London, Hombury St Mary, 
Dorking, RH5 6NT, UK\\}

\date{}

\pagerange{\pageref{firstpage}--\pageref{lastpage}} \pubyear{}

\begin{document}

\maketitle

\label{firstpage}
\begin{abstract}
We investigate radial and vertical metallicity gradients for a sample
of red clump stars from the RAdial Velocity Experiment (RAVE) Data
Release 3. We select a total of 6781 stars, using a selection of
colour, surface gravity and uncertainty in the derived space motion, and
calculate for each star a probabilistic (kinematic) population
assignment to a thin or thick disc using space motion and additionally
another (dynamical) assignment using stellar vertical orbital
eccentricity. We derive almost equal metallicity gradients as a
function of Galactocentric distance for the high probability thin 
disc stars and for stars with vertical orbital eccentricities consistent
with being dynamically young, $e_{v}\leq0.07$,
i.e. $d[M/H]/dR_{m}=-0.041\pm0.003$ and $d[M/H]/dR_{m}=-0.041\pm0.007$
dex kpc$^{-1}$. Metallicity gradients as a function of distance from
the Galactic plane for the same populations are steeper,
i.e. $d[M/H]/dz_{max}=-0.109\pm0.008$ and
$d[M/H]/dz_{max}=-0.260\pm0.031$ dex kpc$^{-1}$, respectively. 
$R_{m}$ and $z_{max}$ are the arithmetic mean of the perigalactic 
and apogalactic distances, and the maximum distance to the Galactic 
plane, respectively. Samples including more thick disc red clump giant 
stars show systematically shallower abundance gradients. These findings 
can be used to distinguish between different formation scenarios of the 
thick and thin discs.
\end{abstract}


\section{Introduction}
Metallicity gradients play an important role in understanding the formation of 
disc populations of the galaxies. In the Milky Way Galaxy, there is extensive 
information establishing a radial gradient in young stars and in the 
interstellar medium \citep*{Shaver83, Luck06, Luck11b}. Values typically 
derived are $d[Fe/H]/dR_{G}=-0.06\pm0.01$ dex kpc$^{-1}$, within 2-3 kpc of 
the Sun. Much effort to search for local abundance variations, and abundance 
variations with azimuth, show little if any detectable variation in young 
systems \citep{Luck11a}, showing the inter-stellar medium to be
well-mixed on quite large scales, supporting the importance of gas flows. 

Extant data suggest vertical metallicity gradients in the
 $-0.4<d[M/H]/dz<-0.2$ dex kpc$^{-1}$ range for relatively small distances 
from the Galactic plane, i.e. $z<4$ kpc \citep*{Trefzger95,Karaali03, Du04, Ak07a,Peng11}. 
For intermediate $z$ distances, where the thick disc is dominant, the vertical 
metallicity gradient is low, $d[M/H]/dz=-0.07$ dex kpc$^{-1}$ and the radial 
gradient is only marginal, $-0.02\leq d[M/H]/dR\leq0$ dex kpc$^{-1}$ \citep{Rong01}. 
There is some evidence that metallicity gradients for relatively short vertical 
distances, $z<2.5$ kpc, show systematic fluctuations with Galactic longitude, 
similar to those of thick-disc scaleheight, which may be interpreted as a common 
flare effect of the disc \citep{Bilir06, Cabrera-Lavers07, Ak07b, Bilir08, Yaz10}.  

Quantifying the abundance distribution functions and their radial and
vertical gradients in both thin and thick discs can be achieved using
stellar abundances, especially those from major surveys such as RAdial
Velocity Experiment \citep[RAVE;][]{Steinmetz06} and Sloan Digital Sky
Survey \citep[SDSS;][]{Abazajian03}.  RAVE is a multi-fibre
spectroscopic astronomical survey in the Milky Way, which covers just
over the half of the Southern hemisphere, using the 1.2-m UK Schmidt
Telescope of the Australian Astronomical Observatory
\citep{Steinmetz06, Zwitter08, Siebert11}.  RAVE's primary aim is to
derive the radial velocity of stars from the observed spectra for
solar neighbourhood stars. Additional information is also derived,
such as photometric parallax and stellar atmospheric parameters,
i.e. effective temperature, surface gravity, metallicity and elemental
abundance data. This information is important in calculating
metallicity gradients, which provides data about the formation and
evolution of the Galaxy. As the data were obtained from solar
neighbourhood stars, we have limitations to distance and range of
metallicity. However, the metallicity measurements are of high
internal accuracy which is an advantage for our work.

In a recent study carried out by \citet{Chen11}, a vertical
metallicity gradient of -0.22 dex kpc$^{-1}$ was claimed for the thick
disc.  They used the SDSS DR8 \citep{Aihara11} data set to identify a
sample of red horizontal branch stars (RHB) and for this sample they
derived the steepest metallicity gradient for the thick disc currently 
in the literature. RHB stars are very old, on average, so it is feasible 
that they can have a different metallicity variation than younger stars. 
However, the difference between their metallicity gradient and others 
in the literature for the thick disc is large, which motivates 
confirmation by other works with different data. 

In our previous study \citep[][paper II]{Coskunoglu11a}, we
investigated the metallicity gradient of a dwarf sample and we
confirmed the radial metallicity gradient of -0.04 dex kpc$^{-1}$
based on calibrated metallicities from RAVE DR3 \citep[][and the references 
therein, Paper I]{Coskunoglu11a}. Additionally, we showed that the radial metallicity 
gradient is steeper for our sample which is statistically selected to favour 
younger stars, i.e. F-type stars with orbital eccentricities $e_{V}\leq0.04$,
i.e. $d[M/H]/dR_{m}=-0.051\pm0.005$ dex kpc$^{-1}$. Vertical metallicity 
gradients could not be derived in this study due to short $z$-distances from 
the Galactic plane. Therefore, in this present study, we analyses stellar 
abundance gradients of red clump stars from RAVE data.

Red Clump (RC) stars are core helium-burning stars, in an identical evolutionary 
phase to those which make up the horizontal branch in globular clusters. However, 
in intermediate-and higher-metallicity systems only the red end of the 
distribution is seen, forming a clump of stars in the colour-magnitude 
diagram. In recent years much work has been devoted to studying the 
suitability of RC stars for application as a distance indicator. Their absolute 
magnitude in the optical ranges from $M_{V}=+0.70$ mag for those of spectral 
type G8 III to $M_{V}=+1.0$ mag for type K2 III \citep{Keenan99}.
The absolute magnitude of these stars in the $K$ band is $M_{K}=-1.61\pm0.03$ 
mag with negligible dependence on metallicity \citep{Alves00}, but with a
real dispersion (see below), and in the $I$ band $M_{I}=-0.23\pm0.03$ mag, 
again without dependence on metallicity \citep{Paczynski98}.
RAVE DR3 red clump giants stars occupy a relatively larger $z$-distance 
interval than do RAVE DR3 dwarfs. Hence, we should have sufficient
data to be able to derive abundance gradients for both directions, 
radial and vertical. As the RHB stars are on the extended branch of RC 
stars, we anticipate the results to be similar to RHB stars
claimed in \citet{Chen11}, and so are able to test their result.

The structure of the paper is: Data selection is described in Section 2; calculated 
space velocities and orbits of sample stars is described in Section 3. Population 
analysis and results are given in Section 4 and Section 5, respectively. Finally, 
a discussion and conclusion are presented in Section 6.
          
\section{Data}
The data were selected from the third data release of RAVE
 \citep[DR3;][]{Siebert11}. RAVE DR3 reports 83072 radial velocity measurements 
for stars $9\leq I\leq12$ mag. This release also provides stellar atmospheric 
parameters for 41 672 spectra representing 39833 individual stars 
\citep{Siebert11}. The accuracy of the radial velocities is high, marginally 
improved with DR3: the distribution of internal errors in radial velocities 
has a mode of 0.8 km s$^{-1}$ and a median of 1.2 km s$^{-1}$, while 95 per 
cent of the sample has an internal error smaller than 5 km s$^{-1}$. The 
uncertainties for the stellar atmospheric parameters are typically: 250 K for 
effective temperature $T_{eff}$, 0.43 dex for surface gravity $\log g$ and 
0.2 dex for $[M/H]$. While RAVE supports a variety of chemical abundance scales, 
we use here just the public DR3 values. Since anticipated gradients are small, 
this provides a well-defined set of parameters for analysis. The proper motions 
of the stars were taken from RAVE DR3 values, which were compiled from PPMX, 
{\it Tycho-2}, SSS and UCAC2 catalogs. The distribution of RAVE DR3 stars in 
the Equatorial and Galactic coordinate planes are shown in Fig. 1. 

\begin{figure}
\begin{center}
\includegraphics[scale=0.50, angle=0]{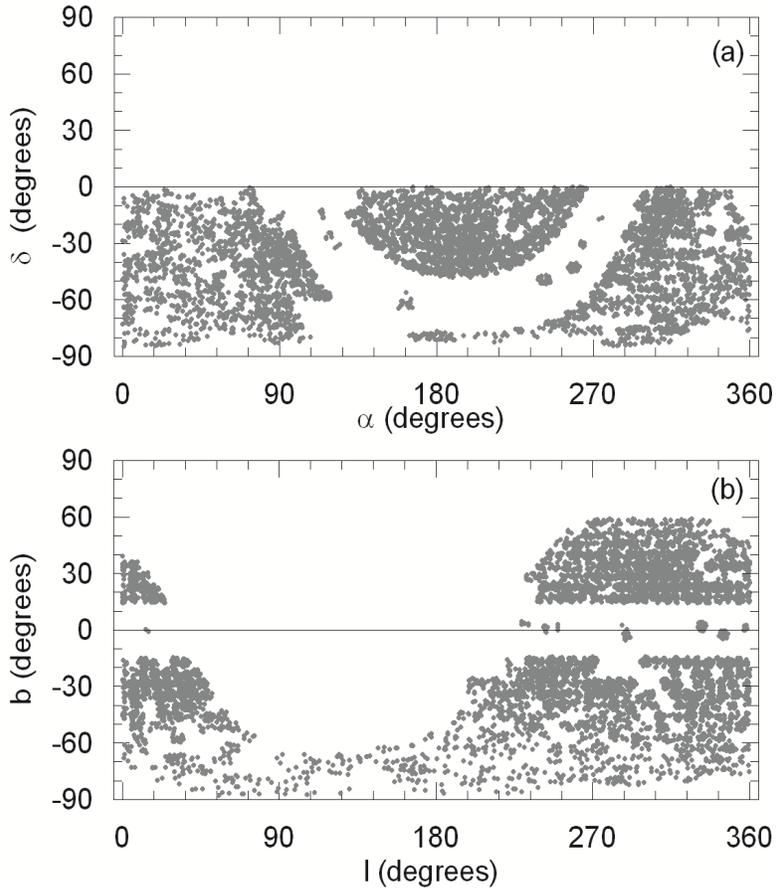}
\caption[] {Distribution of RAVE DR3 stars in the Equatorial (top) and Galactic (bottom) 
coordinate planes.}
\end{center}
\end{figure}

We applied the following constraints to obtain a homogeneous RC sample with 
best quality: i) $2\leq\log g~(cm~s^{-2})\leq3$ \citep{Puzeras10}, ii) the Two Micron 
All Sky Survey \citep[2MASS;][]{Skrutskie06} photometric data are of quality 
labelled as ``AAA'', and iii) ´$(J-H)_{0}>0.4$ \citep{Bilir11}. The numerical values for 
$\chi^2$ and the median of S/N value of sample spectra (totally 7985 stars), thus 
obtained are 671 and 41, respectively. Proper motions for 139 out of the 
7985 stars are not available in the RAVE DR3, hence these values are provided 
from the PPMXL Catalog of \citet*{Roeser10}. Distances were obtained by 
combining the apparent $K_s$ magnitude of the star in query and the absolute 
magnitude $M_{{K}_S}=-1.54\pm0.04$ mag, adopted for all RC stars 
\citep{Groenewegen08}. Whereas the $E(B-V)$ reddening were obtained 
iteratively by using published methodology \citep[for more detailed 
information regarding the iterations see][and the references therein;]
{Coskunoglu11b}. Then, the de-reddening of magnitudes 
and colours in 2MASS were carried out by the following equations with 
the co-efficient of \citet{Fiorucci03}:

\begin{eqnarray}
J_{o}=J-0.887\times E(B-V) \nonumber \\
(J-H)_{o}= (J-H)-0.322\times E(B-V)\\ \nonumber
(H-K_{s})_{o}=(H-K_{s})-0.183\times E(B-V)\nonumber
\end{eqnarray}

Note the real dispersion in absolute magnitude among RC stars (Fig. 2). This 
will have the effect of blurring the derived distances, and so smoothing any 
derived gradient. Given that we search for a linear gradient, such distance 
uncertainties will tend to somewhat reduce any measured gradient. Given our 
results below, we consider any such effect to be second order.

\begin{figure}
\begin{center}
\includegraphics[scale=0.50, angle=0]{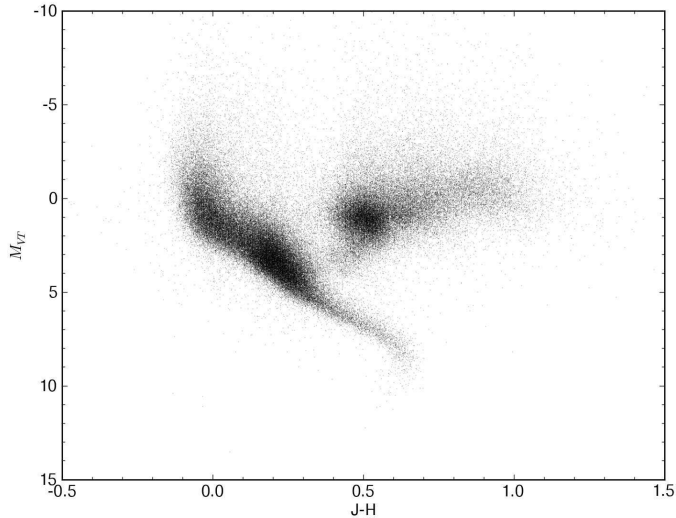}
\caption[] {The colour-absolute magnitude diagram for the revised Hipparcos 
catalogue \citep{vanLeeuwen2007} cross-matched with 2MASS \citep{Cu03}. 
The very well illustrated red clump justifies the colour selection used 
in our study.}
\end{center}
\end{figure}

The distance range of the sample and the median of the distances are
$0.2\leq d\leq3.4$ kpc and 1.34 kpc, respectively (Fig. 3), which are
sufficient to investigate both radial and vertical metallicity
gradients.  The distribution of colour excess $E(B-V)$ is given in
three categories, i.e.  $0^{\circ}<|b|\leq30^{\circ}$,
$30^{\circ}<|b|\leq60^{\circ}$, and $60^{\circ}<|b|\leq90^{\circ}$,
whose mean values are 0.14, 0.06, and 0.02 mag, respectively
(Fig. 4). The reddening is rather small at intermediate and high
Galactic latitudes, as expected. The projection of the sample stars
onto the ($X$, $Y$) and ($X$, $Z$) planes (Fig. 5) show that their
distribution is biassed (by design: RAVE does not observe towards the 
Galactic bulge and so), the median of the heliocentric coordinates are 
$X=0.52$, $Y=-0.59$, $Z=-0.41$ kpc.

\begin{figure}
\begin{center}
\includegraphics[scale=0.45, angle=0]{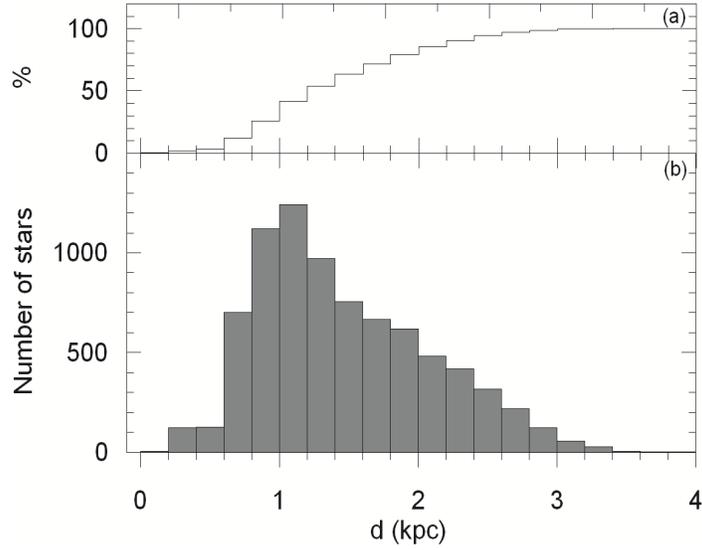}
\caption[] {Cumulative (panel a) and frequency (panel b) distributions of
distances of our sample of RAVE red clump giants.}
\end{center}
\end{figure}

\begin{figure}
\begin{center}
\includegraphics[scale=0.45, angle=0]{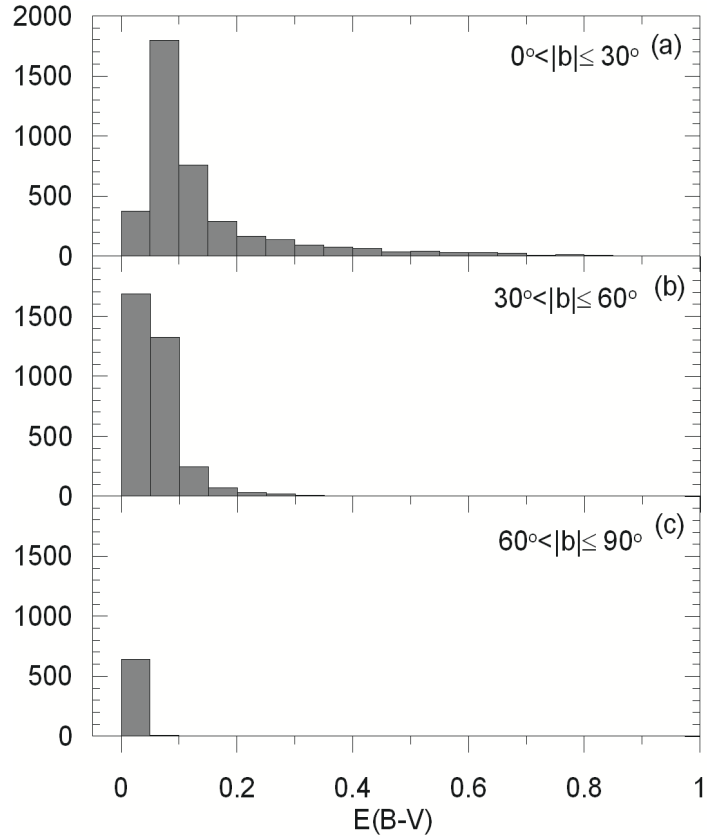}
\caption[] {Distribution of colour excess $E(B-V)$ for our sample of
  RAVE red clump giants.}
\end{center}
\end{figure}

\begin{figure}
\begin{center}
\includegraphics[scale=0.50, angle=0]{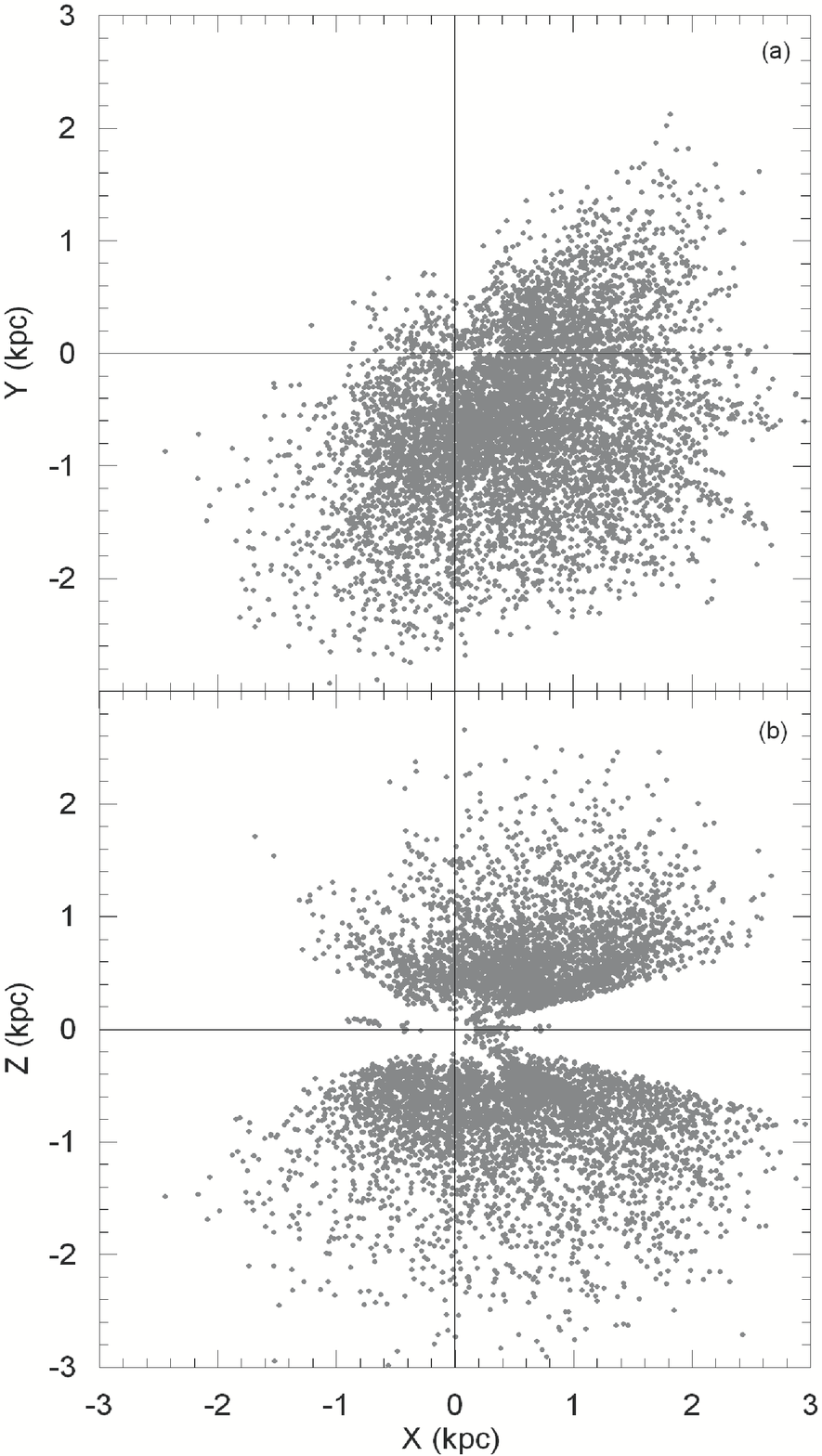}
\caption[] {Space distributions of our sample of RAVE red clump giants
  projected onto two Galactic planes: (a) X-Y and (b) X-Z.}
\end{center}
\end{figure}

\section{Space Velocities and Orbits}
We combined the distances estimated in Section 2 with RAVE kinematics
and the available proper motions, applying the algorithms and the 
transformation matrices of \cite{Johnson87} to obtain their Galactic space 
velocity components ($U$, $V$, $W$). In the calculations epoch J2000 was 
adopted as described in the International Celestial Reference System (ICRS) 
of the {\em Hipparcos} and {\em Tycho-2} Catalogues \citep{ESA97}. The 
transformation matrices use the notation of a right handed system. Hence, 
$U$, $V$, and $W$ are the components of a velocity vector of a star with 
respect to the Sun, where $U$ is positive towards the Galactic centre
($l=0^{\circ}$, $b=0^{\circ}$), $V$ is positive in the direction of
Galactic rotation ($l=90^{\circ}$, $b=0^{\circ}$) and $W$ is positive
towards the North Galactic Pole ($b=90^{\circ}$).

We adopted the value of the rotation speed of the Sun as 222.5 km s$^{-1}$.
Correction for differential Galactic rotation is necessary for
accurate determination of $U$, $V$, and $W$ velocity components. The
effect is proportional to the projection of the distance to the stars
onto the Galactic plane, i.e. the $W$ velocity component is not
affected by Galactic differential rotation \citep{Mihalas81}. We
applied the procedure of \cite{Mihalas81} to the distribution of the
sample stars and estimated the first order Galactic differential
rotation corrections for $U$ and $V$ velocity components of the sample
stars. The range of these corrections is $-92<dU<58$ and
$6<dV<7$ km s$^{-1}$ for $U$ and $V$, respectively. As
expected, $U$ is affected more than the $V$ component. Also, the high
values for the $U$ component show that corrections for differential
Galactic rotation can not be ignored. One notices that Galactic differential 
rotation corrections are rather larger than the corresponding ones for dwarfs 
\citet{Coskunoglu11b}. The $U$, $V$, and $W$ velocities were reduced to LSR 
by adopting the solar LSR velocities in \citep{Coskunoglu11b}, i.e. 
($U_{\odot}$, $V_{\odot}$, $W_{\odot}$)=(8.83, 14.19, 6.57) km s$^{-1}$. 

The uncertainty of the space velocity components $U_{err}$, $V_{err}$
and $W_{err}$ were computed by propagating the uncertainties of the
proper motions, distances and radial velocities, again using an 
algorithm by \cite{Johnson87}. Then, the error for the
total space motion of a star follows from the equation:
\begin{equation}
S_{err}^{2}=U_{err}^{2}+V_{err}^{2}+W_{err}^{2}.
\end{equation}
The mean S$_{err}$ and standard deviation ($s$) for space velocity
errors are S$_{err}=39$ km s$^{-1}$ and $s=36$ km s$^{-1}$,
respectively. We now remove the most discrepant data from the
analysis, knowing that outliers in a survey such as this will
preferentially include stars which are systematically mis-analysed
binaries, etc. Astrophysical parameters for such stars are also likely
to be relatively unreliable. Thus, we omit stars with errors that
deviate by more than the sum of the standard error and the standard
deviation, i.e. $S_{err}>$ 75 km s$^{-1}$. This removes 1204 stars,
$\sim$15.1 per cent of the sample. Thus, our sample was reduced to
6781 stars, those with more robust space velocity components. After
applying this constraint, the mean values and the standard deviations
for the velocity components were reduced to ($U_{err}$, $V_{err}$,
$W_{err}$)=($15.03\pm10.61$, $15.12\pm11.20$, $15.68\pm12.06$) km
s$^{-1}$.  The distribution of the errors for the space velocity
components is given in Fig. 6. In this study, we used only the
sub-sample of stars (6781 stars) with standard error $S_{err} \leq75$
km s$^{-1}$. The $U,V,W$ velocity diagrams for these stars are shown
in Fig. 7. The centre of the distributions deviate from the zero
points of the $U$, $V$, and $W$ velocity components, further indicating the
need for a Local Standard of Rest reduction.

\begin{figure}
\begin{center}
\includegraphics[scale=0.70, angle=0]{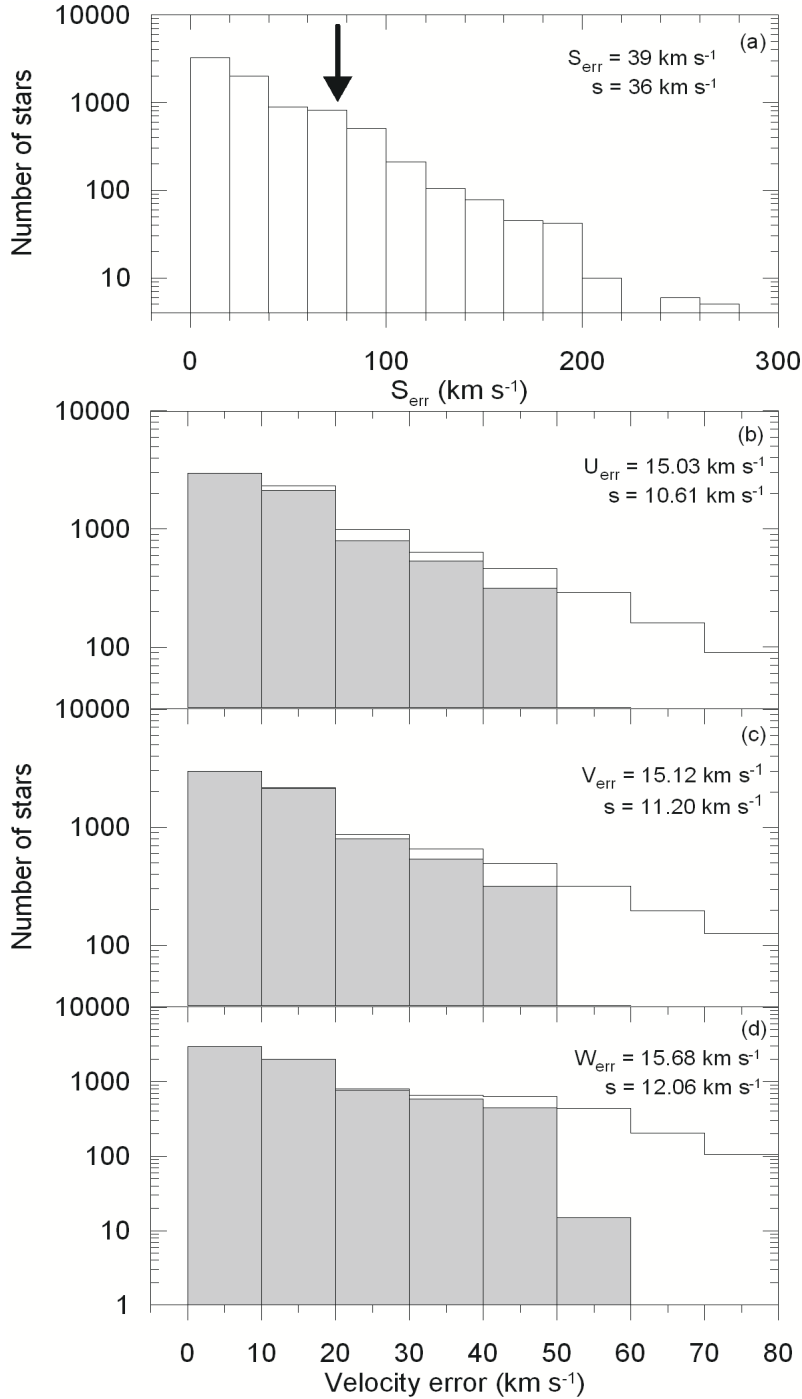}
\caption[] {Error histograms for space velocity (panel a) and its components
(panels b-d) for our sample of RAVE red clump stars. The arrow in panel (a) indicates
the upper limit of the total error adopted in this work. The shaded part of
the histogram indicates the error for different velocity components of stars
after removing the stars with large space velocity errors.}
\end{center}
\end{figure}

\begin{figure}
\begin{center}
\includegraphics[scale=0.50, angle=0]{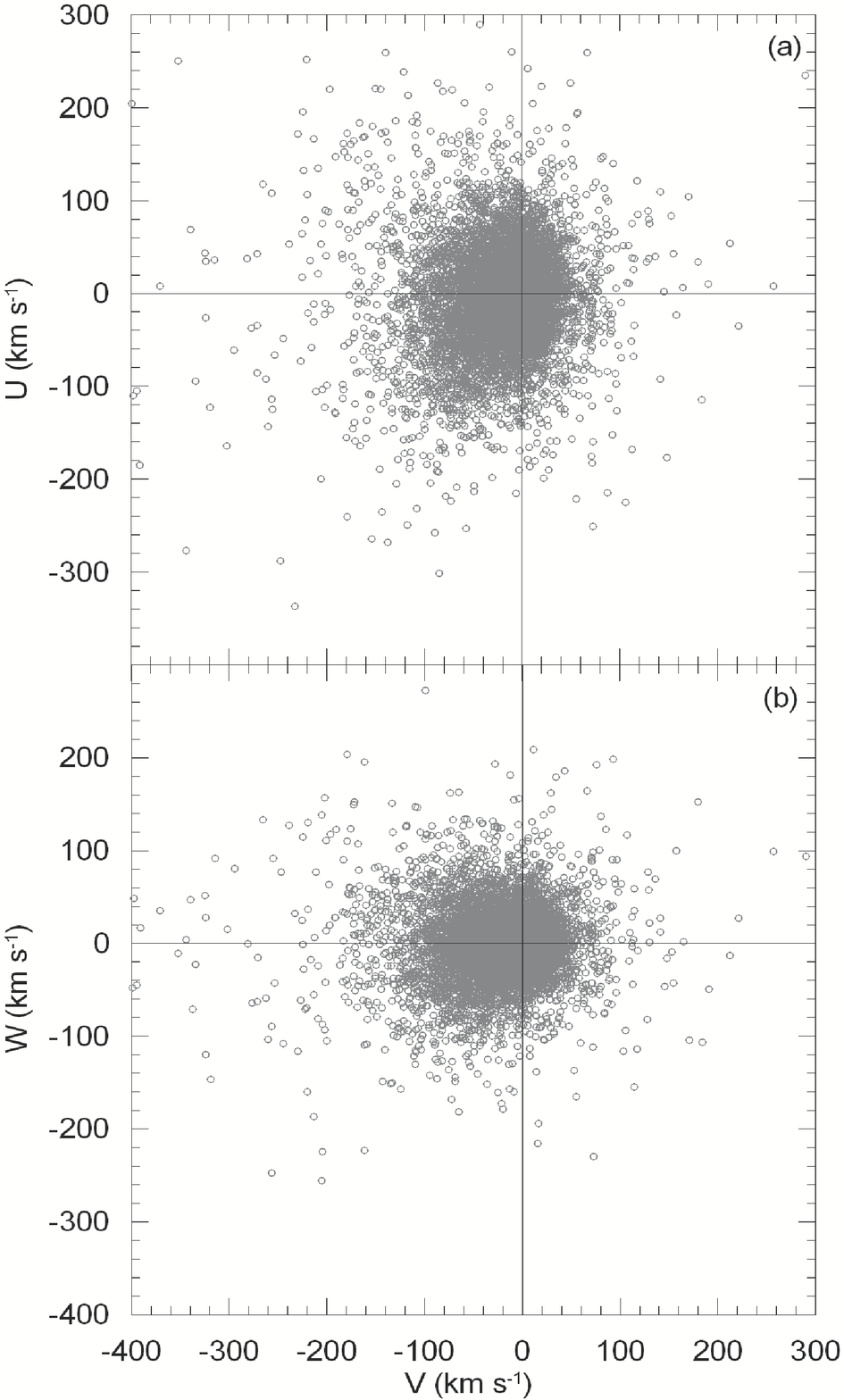}
\caption[] {The distribution of velocity components of our final
  cleaned sample of RAVE red clump stars with high-quality data, in
  two projections: (a) $U-V$ and (b) $W-V$.}
\end{center}
\end{figure}

To complement the chemical abundance data, accurate kinematic data
have been obtained and used to calculate individual Galactic orbital
parameters for all stars. The shape of the stellar orbit is a proxy,
through the age-velocity dispersion relation, for age, with more
circular orbits hosting statistically younger stars. 

In order to calculate those parameters we used standard gravitational 
potentials well-described in the literature \citep*{Miyamoto75, 
Hernquist90, Johnston95, Dinescu99} to estimate orbital elements of 
each of the sample stars. The orbital elements for a star used in our 
work are the mean of the corresponding orbital elements calculated over 
15 orbital periods of that specific star. The orbital integration 
typically corresponds to 3 Gyr, and is sufficient to evaluate the 
orbital elements of solar neighbourhood stars, most of which have 
orbital periods below 250 Myr.

Solar neighbourhood velocity space includes well-established
substructures that resemble classic moving groups or stellar streams
\citep*{Dehnen98, Skuljan99, Nordstrom04}. \citet{Famaey05}, \citet*{Famaey08} 
and \cite{Pompeia11} show that, although these streams include clusters,
after which they are named, and evaporated remnants from these
clusters, the majority of stars in these streams are not coeval but
include stars of different ages, not necessarily born in the same
place nor at the same time. They argue these streams are dynamical
(resonant) in origin, probably related to dynamical perturbations by
transient spiral waves \citep*{DeSimone04}, which migrate
stars to specific regions of the $UV$-plane. Stars in a dynamical
stream just share a common velocity vector at this particular epoch.
These authors further point out the obvious and important point that
dynamical streams are kinematically young and so integrating backwards
in a smooth stationary axisymmetric potential the orbits of the stars
belonging to these streams is non-physical. RAVE stars are selected to 
avoid the Galactic plane ($|b|>10^{\circ}$) . Dynamical perturbations 
by transient spiral waves are strongest closest to the Galactic plane 
so there will be fewer dynamical stream stars in our RAVE sample. Hence, 
contamination of the dynamical stream stars is unlikely to affect our
statistical results.    

To determine a possible orbit, we first perform test-particle
integration in a Milky Way potential which consists of a logarithmic
halo of the form
\begin{eqnarray}
  \Phi_{\rm halo}(r)=v_{0}^{2} \ln \left(1+\frac{r^2}{d^2}\right),
\end{eqnarray}
with $v_{0}=186$ km s$^{-1}$ and $d=12$ kpc. The disc is represented
by a Miyamoto-Nagai potential:
\begin{eqnarray}
  \Phi_{\rm disc}(R,z)=-\frac{G M_{\rm d}} { \sqrt{R^{2} + \left(
        a_d + \sqrt{z^{2}+b_d^{2}} \right)^{2}}},
\end{eqnarray}
with $M_{\rm d}=10^{11}~M_{\odot}$, $a_d=6.5$ kpc and $b_d=0.26$
kpc. Finally, the bulge is modelled as a Hernquist potential
\begin{eqnarray}
  \Phi_{\rm bulge}(r)=-\frac{G M_{\rm b}} {r+c},
\end{eqnarray}
using $M_{\rm b}=3.4\times10^{10}~M_{\odot}$ and $c=0.7$ kpc.  The
superposition of these components gives quite a good representation of
the Milky Way. The circular speed at the solar radius is $\sim 220$ km
s$^{-1}$. $P_{LSR}=2.18\times10^8$ years is the orbital period of the
LSR and $V_c=222.5$ km s$^{-1}$ denotes the circular rotational
velocity at the solar Galactocentric distance, $R_0=8$ kpc.

For our analysis of gradients, we are interested in the mean radial
Galactocentric distance ($R_m$) as a function of the stellar
population and the orbital shape. \cite{Wilson11} has analysed the
radial orbital eccentricities of a RAVE sample of thick disc stars, to
test thick disc formation models. Here we focus on possible local
gradients, so instead consider the {\it vertical} orbital
eccentricity, $e_V$. $R_m$ is defined as the arithmetic mean of the
final perigalactic ($R_p$) and apogalactic ($R_a$) distances, 
and $z_{max}$ and $z_{min}$ are the final maximum and minimum distances, 
respectively, to the Galactic plane. Whereas $e_v$ is defined as follows:

\begin{eqnarray}
e_v=\frac{(|z_{max}|+|z_{min}|)}{R_m},
\end{eqnarray}
where $R_m=(R_a+R_p)/2$ \citep{Pauli05}. Due to $z$-excursions
$R_p$ and $R_a$ can vary, however this variation is not more than
5 per cent.

\section{Population Analysis}
\subsection{Classification using space motions}
The procedure of \cite*{Bensby03, Bensby05} was used to separate sample
stars into different populations. This kinematic methodology assumes that
Galactic space velocities for the thin disc ($D$), thick disc ($TD$),
and stellar halo ($H$) with respect to the LSR have Gaussian
distributions as follows:

\begin{equation}
f(U,~V,~W)=k~\times~\exp\Biggl(-\frac{U_{LSR}^{2}}{2\sigma_{U{_{LSR}}}^{2}}-\frac{(V_{LSR}-V_{asym})
^{2}}{2\sigma_{V{_{LSR}}}^{2}}-\frac{W_{LSR}^{2}}{2\sigma_{W{_{LSR}}}^{2}}\Biggr),
\end{equation}
where
\begin{equation}
k=\frac{1}{(2\pi)^{3/2}\sigma_{U{_{LSR}}}\sigma_{V{_{LSR}}}\sigma_{W{_{LSR}}}},
\end{equation}
normalizes the expression. For consistency with other analyses
$\sigma_{U{_{LSR}}}$, $\sigma_{V{_{LSR}}}$ and $\sigma_{W{_{LSR}}}$
were adopted as the characteristic velocity dispersions: 35, 20 and 16
km s$^{-1}$ for thin disc ($D$); 67, 38 and 35 km s$^{-1}$ for thick
disc ($TD$); 160, 90 and 90 km s$^{-1}$ for halo ($H$), respectively
\citep{Bensby03}. $V_{asym}$ is the asymmetric drift: -15, -46 and
-220 km s$^{-1}$ for thin disc, thick disc and halo, respectively. LSR
velocities were taken from \cite{Coskunoglu11b} and these values are
$(U, V, W)_{LSR}=(8.83\pm0.24, 14.19\pm0.34, 6.57\pm0.21)$ km
s$^{-1}$.

The probability of a star of being ``a member'' of a given population 
with respect to a second population is defined as the ratio of the 
$f(U, V, W)$ distribution functions times the ratio of the local space 
densities for two populations. Thus,

\begin{equation}
TD/D=\frac{X_{TD}}{X_{D}}\times\frac{f_{TD}}{f_{D}}~~~~~~~~~~TD/H=\frac{X_{TD}}{X_{H}}\times\frac{f_{TD}}{f_{H}},
\end{equation}
are the probabilities for a star being classified as a thick disc star
relative to it being a thin disc star, and relative to it being a halo
star, respectively. $X_{D}$, $X_{TD}$ and $X_{H}$ are the local space
densities for thin disc, thick disc and halo, i.e. 0.94, 0.06, and
0.0015, respectively \citep*{Robin96, Buser99}. We followed the
argument of \cite{Bensby05} and separated the sample stars into four
categories: $TD/D\leq 0.1$ (high probability thin disc stars),
$0.1<TD/D\leq 1$ (low probability thin disc stars), $1<TD/D\leq 10$
(low probability thick disc stars) and $TD/D>10$ (high probability
thick disc stars). Fig. 8 shows the $U-V$ and $W-V$ diagrams as a
function of population types defined by using \cite{Bensby03}'s criteria. 
It is evident from Fig. 8 that the kinematic population assignments are 
strongly affected by space-motion uncertainties. 3385 and 1151 stars of the 
sample were classified as high and low probability thin disc stars, 
respectively, whereas 646 and 1599 stars are low and high 
probability thick disc stars (Table 1). The relative number of high 
probability thick disc (RC) stars are much larger than the corresponding 
ones in \cite{Coskunoglu11a} (2 per cent), i.e. 24 per cent.   

\begin{figure}
\begin{center}
\includegraphics[scale=0.70, angle=0]{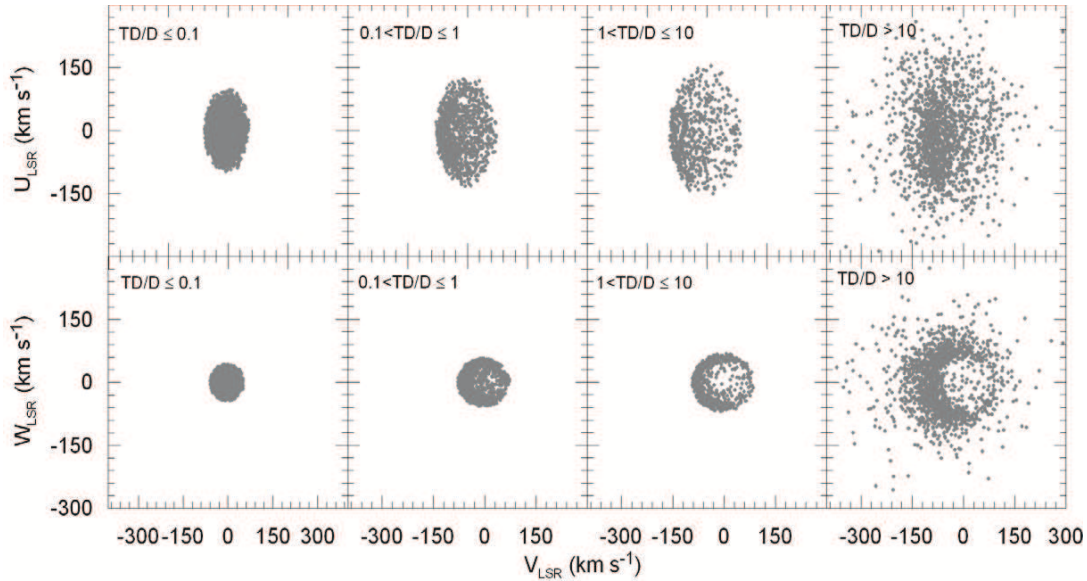}
\caption[] {$U-V$ and $W-V$ diagrams for our sample of red clump stars, applying
  \citet{Bensby03}'s population classification criterion. It is
  apparent that space-motion uncertainities remain significant, even
  for this sample.}
\end{center}
\end{figure}

\begin{table}
\center 
\caption{The space velocity component ranges for the range of
  population types into which our RAVE red clump stars have been classified.
}
\begin{tabular}{ccccc}
\hline
Parameters &  $U$ & $V$ & $W$ & N\\
     &  (km s$^{-1}$) & (km s$^{-1}$) & (km s$^{-1}$) & \\
\hline
$TD/D \leq 0.1$     &  (-98, 97)   & (-60, 51)  & (-44, 43)  & 3385 \\
$0.1 < TD/D \leq 1$ &  (-133, 123) & (-78, 72)  & (-57, 58)  & 1151 \\
$1 < TD/D \leq 10$  &  (-151, 154) & (-93, 85)  & (-69, 69)  & 646  \\
$TD/D > 10$         &  (-337, 290) & (-400, 291)& (-344, 273)& 1599 \\
\hline
\end{tabular}  
\end{table}

\subsection{Population classification using stellar vertical orbital shape}
Both radial and vertical orbital eccentricities contain valuable
information: here we consider the vertical orbit shape. Vertical
orbital eccentricities were calculated, as described above, from
numerically-integrated orbits. We term this the dynamical method of
population assignment, which complements the \cite{Bensby03}'s approach.
The distribution function of $e_{V}$ is not consistent with a single Gaussian 
distribution, as is shown in Fig. 9. A two-Gaussian model however 
does provide an acceptable fit. For convenience, we separated our sample 
into three categories, i.e. stars with $e_v\leq0.12$ 
(3448 stars), $0.12<e_v\leq0.25$ (2389 stars) and $e_v>0.25$ 
(944 stars), and fitted their metallicities to their mean 
radial distances ($R_{m}$) in order to investigate the 
presence of a metallicity gradient for RAVE RC stars. We 
provide in Table 2 (electronically) for each star, stellar 
parameters from RAVE DR3, calculated kinematical and dynamical
parameters and our stellar population assignment.

\begin{figure*}
\begin{center}
\includegraphics[scale=0.70, angle=0]{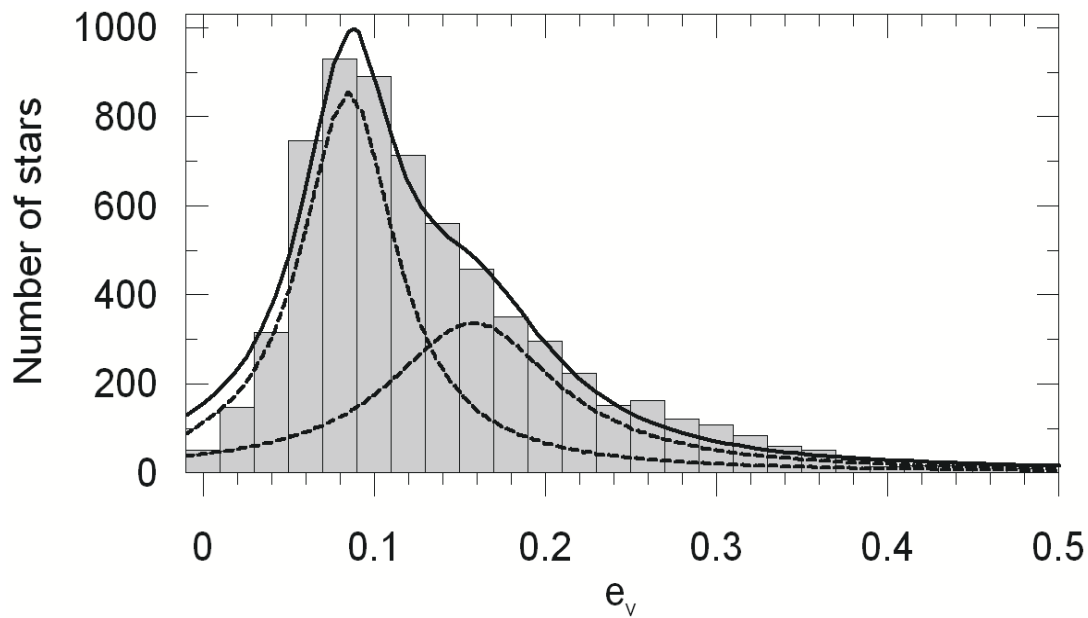}
\caption[] {The distribution of vertical orbital eccentricity for our
  sample of RC stars. The dashed lines indicate to the stars with $e_V\leq0.12$ (the one with higher mod) and $0.12<e_V\leq0.25$ eccentricities (lower mod), respectively, whereas the solid line corresponds to the whole sample.}
\end{center}
\end{figure*}

\begin{landscape}
\begin{table*}
\setlength{\tabcolsep}{4pt}
{\scriptsize
\center
\caption{Stellar atmospheric parameters, astrometric, kinematic and dynamic data for the whole sample: (1): Our catalogue number, (2): RAVEID, (3-4): Equatorial coordinates in degrees (J2000), (5): $T_{\textrm{eff}}$ in K, (6) $\log g$ (cm~s$^{-2}$) in dex, (7): Calibrated metallicity $[M/H]$ (dex), (8-9): Proper motion components in mas/yr, (10): $d$ in kpc, (11): Heliocentric radial velocity in km s$^{-1}$, (12-17): Galactic space velocity components, and their respective errors in km s$^{-1}$, (18-19): Perigalactic and apogalactic distance in kpc, (20-21): Minimum and maximum distance from the Galactic plane in kpc, (22): $TD/D$ ratio as mentioned in the text. $TD/D=999.999$ shows the value of $TD/D>999$.}
\begin{tabular}{lccccccccccccccccccccc}
\hline
(1)	&(2)	&(3)	&(4)	&(5)	&(6)	&(7)	&(8)	&(9)	&(10)	&(11)	&(12)	&(13)	&(14)	&(15)	&(16)	&(17)	&(18)	&(19)	&(20)	&(21)	&(22)\\
ID	& Designation	& $\alpha$ & $\delta$	& $T_{eff}$	& $\log g$	& $[M/H]$	& $\mu_\alpha \cos \delta$	& $\mu_\delta$	& $d$	& $\gamma$	& $U_{LSR}$	& $U_{err}$	& $V_{LSR}$	& $V_{err}$	& $W_{LSR}$	& $W_{err}$ & $d_{per}$	& $d_{apo}$	& $z_{min}$	& $z_{max}$	& $TD/D$\\
\hline
 1 & J000002.9-490434 &   0.012167 & -49.076083 &       4963 &       2.99 &      -0.32 &       20.2 &      -11.7 &      1.920 &     -25.23 &    -109.28 &      20.12 &    -154.86 &      21.00 &      31.23 &       8.61 &      1.822 &      8.580 &     -2.096 &      4.316 &    999.999 \\
 2 & J000013.7-725608 &   0.056875 & -72.935667 &       4491 &       2.24 &       0.01 &       18.8 &       -4.0 &      0.885 &      10.40 &     -38.44 &      13.36 &     -39.40 &      11.97 &      -4.49 &      10.36 &      5.433 &      7.994 &     -0.627 &      0.628 &      0.027 \\
 3 & J000019.8-283022 &   0.082458 & -28.505972 &       4596 &       2.14 &      -0.43 &       19.2 &       -9.2 &      1.243 &      18.10 &     -64.75 &      16.37 &     -83.65 &      16.70 &     -32.70 &       3.37 &      3.582 &      8.415 &     -1.397 &      1.398 &     32.168 \\
 4 & J000207.7-540333 &   0.532167 & -54.059028 &       4588 &       2.06 &      -0.54 &       15.5 &      -10.0 &      1.631 &      14.53 &     -50.67 &      23.08 &    -110.26 &      22.61 &       4.76 &      11.20 &      2.790 &      7.880 &     -1.444 &      1.444 &    359.507 \\
 5 & J000208.9-843734 &   0.536875 & -84.626139 &       4648 &       2.21 &      -0.59 &      -11.2 &        2.4 &      0.901 &      84.39 &     106.73 &      10.02 &     -18.51 &       7.93 &     -37.96 &       9.05 &      5.008 &     10.153 &     -0.937 &      0.940 &      1.708 \\
 6 & J000211.9-633612 &   0.549750 & -63.603389 &       4698 &       2.59 &      -0.48 &       13.1 &       -8.3 &      1.449 &      61.38 &     -11.57 &      22.70 &     -99.34 &      22.22 &     -26.45 &      15.03 &      3.129 &      7.538 &     -1.253 &      1.256 &     66.785 \\
 7 & J000214.2-083408 &   0.559042 &  -8.568889 &       5237 &       2.80 &      -0.53 &       13.2 &       -2.6 &      1.199 &      11.97 &     -61.50 &      12.23 &     -27.10 &      11.59 &     -23.51 &       4.74 &      5.805 &      9.049 &     -1.256 &      1.253 &      0.061 \\
 8 & J000241.8-355010 &   0.673958 & -35.836028 &       4661 &       2.86 &      -0.23 &       22.8 &       -6.4 &      1.382 &     -14.50 &    -105.23 &      10.45 &     -90.62 &      11.37 &      -1.53 &       2.63 &      3.177 &      9.171 &     -1.450 &      1.450 &    137.423 \\
 9 & J000257.3-060701 &   0.738542 &  -6.117028 &       4739 &       2.57 &      -0.40 &       29.7 &        7.5 &      1.149 &       1.42 &    -165.05 &      17.71 &     -25.73 &      16.15 &     -10.23 &       7.25 &      4.457 &     13.125 &     -1.446 &      1.462 &     30.648 \\
10 & J000324.5-821243 &   0.852167 & -82.212028 &       4733 &       2.60 &      -0.50 &       -4.2 &      -10.5 &      1.015 &      45.37 &      73.12 &      12.40 &     -37.92 &      10.18 &      24.92 &      11.39 &      4.780 &      8.590 &     -0.725 &      0.726 &      0.193 \\
11 & J000324.5-821243 &   0.852167 & -82.212028 &       4651 &       2.81 &      -0.43 &       -4.2 &      -10.5 &      1.015 &      44.85 &      72.87 &      12.39 &     -37.57 &      10.18 &      25.22 &      11.39 &      4.785 &      8.595 &     -0.729 &      0.731 &      0.191 \\
12 & J000347.2-475241 &   0.946708 & -47.877917 &       4799 &       2.81 &      -0.60 &       28.2 &       14.6 &      0.785 &      -8.62 &    -100.97 &       8.18 &      14.86 &       7.61 &     -23.24 &       3.27 &      6.397 &     12.017 &     -1.035 &      1.036 &      0.322 \\
13 & J000347.5-832736 &   0.947792 & -83.459861 &       4383 &       2.04 &      -0.51 &       -1.3 &       -3.4 &      1.343 &     -15.75 &      36.35 &      34.21 &      14.93 &      28.23 &      34.38 &      32.38 &      6.877 &      9.373 &     -1.053 &      1.049 &      0.062 \\
14 & J000401.7-211407 &   1.006958 & -21.235167 &       4624 &       2.23 &      -0.57 &       -4.3 &        6.1 &      1.511 &      41.46 &      11.93 &      22.23 &      73.43 &      22.44 &     -23.53 &       4.90 &      7.955 &     15.539 &     -2.557 &      2.579 &      2.480 \\
15 & J000401.7-211407 &   1.006958 & -21.235167 &       4763 &       2.62 &      -0.41 &       -4.3 &        6.1 &      1.511 &      42.90 &      12.09 &      22.23 &      73.69 &      22.44 &     -24.93 &       4.88 &      7.956 &     15.588 &     -2.588 &      2.602 &      2.858 \\
16 & J000405.7-454339 &   1.023875 & -45.727583 &       4497 &       2.35 &       0.11 &       -6.0 &       -9.3 &      0.682 &      16.08 &      45.84 &       6.56 &      -6.79 &       6.69 &       4.27 &       2.61 &      6.569 &      8.910 &     -0.689 &      0.690 &      0.010 \\
17 & J000406.6-774539 &   1.027417 & -77.760917 &       4886 &       2.83 &      -0.06 &       -0.4 &        6.4 &      1.438 &     -57.21 &       2.07 &      12.01 &      77.36 &       8.46 &      10.18 &       7.28 &      7.423 &     14.769 &     -1.495 &      1.494 &      2.074 \\
18 & J000431.7-412250 &   1.132083 & -41.380556 &       4415 &       2.43 &      -0.13 &        6.6 &       -0.8 &      1.500 &      15.43 &     -20.36 &      14.44 &     -15.10 &      14.82 &     -15.32 &       4.53 &      6.944 &      8.011 &     -1.485 &      1.485 &      0.009 \\
19 & J000454.7-571529 &   1.228042 & -57.258111 &       4760 &       2.60 &      -0.34 &        9.5 &      -18.7 &      1.739 &      44.73 &      24.17 &      20.72 &    -164.50 &      21.39 &      29.29 &      11.75 &      1.456 &      7.214 &     -3.777 &      1.918 &    999.999 \\
20 & J000509.7-503556 &   1.290583 & -50.598889 &       4830 &       2.14 &      -0.80 &       -6.5 &       -2.1 &      2.091 &      95.62 &     119.22 &      21.64 &      -1.55 &      21.97 &     -60.53 &       9.73 &      5.674 &     11.893 &     -3.556 &      3.791 &     99.018 \\
21 & J000523.6-195000 &   1.348333 & -19.833417 &       4699 &       2.24 &      -0.62 &       -2.2 &      -14.2 &      0.780 &      15.08 &      38.30 &       8.62 &     -24.74 &       8.72 &     -13.68 &       2.01 &      6.185 &      8.473 &     -0.831 &      0.833 &      0.016 \\
22 & J000547.5-285808 &   1.447875 & -28.969000 &       4760 &       2.35 &      -0.49 &        8.7 &      -12.7 &      0.863 &      17.41 &       3.22 &       7.85 &     -47.47 &       9.51 &     -15.60 &       1.50 &      5.286 &      7.886 &     -0.890 &      0.889 &      0.044 \\
23 & J000610.7-391521 &   1.544458 & -39.255861 &       4965 &       2.69 &      -0.42 &       -0.2 &       -0.6 &      1.323 &      11.92 &      18.03 &      14.16 &       9.47 &      10.57 &      -3.96 &       3.37 &      7.564 &      8.994 &     -1.401 &      1.405 &      0.007 \\
24 & J000632.0-525213 &   1.633292 & -52.870139 &       4558 &       2.10 &      -0.52 &        1.9 &        8.1 &      1.301 &      51.14 &       8.92 &      15.11 &      33.84 &      15.31 &     -62.00 &       7.34 &      7.654 &     11.065 &     -2.203 &      2.203 &      6.763 \\
25 & J000711.2-712952 &   1.796500 & -71.497833 &       4890 &       2.72 &      -0.53 &       29.2 &       -9.1 &      1.420 &       4.98 &    -124.67 &      20.65 &    -122.94 &      18.60 &      11.35 &      15.22 &      2.414 &      9.459 &     -1.141 &      1.139 &    999.999 \\
26 & J000740.7-213330 &   1.919500 & -21.558194 &       4727 &       2.05 &      -0.71 &        0.6 &       -4.0 &      0.738 &      39.03 &      14.19 &       8.44 &       7.53 &       8.56 &     -33.49 &       1.87 &      7.812 &      8.822 &     -0.986 &      0.983 &      0.034 \\
27 & J000740.7-213330 &   1.919500 & -21.558194 &       4924 &       2.59 &      -0.47 &        0.6 &       -4.0 &      0.738 &      40.00 &      14.28 &       8.44 &       7.69 &       8.56 &     -34.43 &       1.79 &      7.814 &      8.841 &     -0.997 &      0.998 &      0.037 \\
28 & J000742.0-413624 &   1.924833 & -41.606667 &       4889 &       2.52 &      -0.61 &        7.2 &       -0.9 &      1.294 &     -26.08 &     -29.22 &      11.91 &      -8.21 &      12.18 &      25.37 &       3.68 &      7.146 &      8.328 &     -1.432 &      1.434 &      0.018 \\
29 & J000743.4-200051 &   1.930792 & -20.014083 &       4709 &       2.38 &      -0.15 &       15.1 &      -11.1 &      0.850 &     -10.49 &     -27.96 &      11.09 &     -54.84 &      11.51 &       0.76 &       2.49 &      4.860 &      8.074 &     -0.841 &      0.834 &      0.072 \\
30 & J000743.4-200051 &   1.930792 & -20.014083 &       4575 &       2.24 &      -0.21 &       15.1 &      -11.1 &      0.850 &      -9.34 &     -27.86 &      11.09 &     -54.62 &      11.51 &      -0.36 &       2.60 &      4.872 &      8.075 &     -0.839 &      0.835 &      0.070 \\
31 & J000752.0-211211 &   1.966458 & -21.203056 &       4554 &       2.17 &      -0.72 &       18.9 &      -30.2 &      0.812 &      -9.65 &      -3.35 &      10.53 &    -122.30 &      11.55 &      -8.88 &       2.34 &      2.298 &      7.926 &     -0.935 &      0.942 &    999.999 \\
32 & J000752.0-211211 &   1.966458 & -21.203056 &       4747 &       2.50 &      -0.45 &       18.9 &      -30.2 &      0.812 &      -8.90 &      -3.28 &      10.53 &    -122.17 &      11.55 &      -9.61 &       2.43 &      2.303 &      7.927 &     -0.944 &      0.929 &    999.999 \\
33 & J000755.3-672223 &   1.980417 & -67.372944 &       4642 &       2.46 &      -0.36 &       15.4 &        2.7 &      1.743 &     -42.89 &    -101.65 &      44.54 &      -8.32 &      42.07 &       3.60 &      31.86 &      5.927 &     10.986 &     -1.737 &      1.737 &      0.114 \\
34 & J000811.4-643552 &   2.047417 & -64.597778 &       4551 &       2.59 &      -0.49 &       10.9 &        0.9 &      0.741 &      28.06 &      -4.72 &       5.94 &     -15.01 &       5.64 &     -23.78 &       4.05 &      6.900 &      7.739 &     -0.678 &      0.678 &      0.014 \\
35 & J000812.3-393058 &   2.051042 & -39.516194 &       4966 &       2.58 &      -0.83 &       24.9 &       -8.0 &      1.278 &     -53.86 &    -111.17 &      15.03 &     -93.27 &      15.40 &      43.25 &       4.09 &      3.225 &      9.304 &     -1.881 &      1.882 &    999.999 \\
36 & J000815.0-404803 &   2.062333 & -40.800944 &       4686 &       2.36 &      -0.76 &       10.0 &       -0.2 &      1.716 &      75.80 &     -36.38 &      39.42 &     -35.44 &      40.35 &     -79.37 &      11.47 &      6.205 &      8.411 &     -2.614 &      2.374 &    324.914 \\
37 & J000846.6-400516 &   2.194083 & -40.087667 &       4857 &       2.88 &      -0.16 &        0.0 &        7.7 &      1.116 &      20.39 &      -0.06 &      15.92 &      46.96 &      16.33 &     -21.79 &       4.49 &      7.793 &     11.916 &     -1.527 &      1.530 &      0.102 \\
38 & J000910.9-392316 &   2.295458 & -39.387722 &       4676 &       2.59 &      -0.42 &        9.3 &      -16.3 &      2.001 &      13.43 &       7.64 &      34.08 &    -165.20 &      35.34 &      10.55 &       9.18 &      1.694 &      7.425 &     -3.592 &      1.602 &    999.999 \\
39 & J000915.5-210409 &   2.314500 & -21.069083 &       4659 &       2.89 &      -0.31 &      -14.0 &       -1.1 &      2.414 &      21.10 &     144.85 &      50.39 &      80.89 &      49.64 &      10.60 &      10.12 &      7.023 &     25.152 &     -5.456 &      6.218 &    999.999 \\
40 & J000942.1-565953 &   2.425375 & -56.997917 &       4895 &       2.18 &      -0.70 &       15.1 &        3.0 &      0.939 &       1.69 &     -44.22 &       8.51 &      -7.76 &       8.30 &     -12.20 &       4.62 &      6.694 &      8.729 &     -0.884 &      0.883 &      0.012 \\
...	&...	&...	&...	&...	&...	&...	&...	&...	&...	&...	&...	&...	&...	&...	&...	&...	&...	&...	&...	&...	&...\\
...	&...	&...	&...	&...	&...	&...	&...	&...	&...	&...	&...	&...	&...	&...	&...	&...	&...	&...	&...	&...	&...\\
...	&...	&...	&...	&...	&...	&...	&...	&...	&...	&...	&...	&...	&...	&...	&...	&...	&...	&...	&...	&...	&...\\
\hline
\end{tabular}
}
\end{table*}
\end{landscape} 

\section{Results}
\subsection{First hints of a metallicity gradient apparent in the RC sample 
of stars.} 
We show in Fig. 10 the distribution of metallicities for our final sample of 
RAVE RC stars. The metallicity distribution for all populations is rather 
symmetric with a mode at $[M/H]\sim-0.4$ dex. In Fig. 11 we show the normalized
metallicity distribution functions, with the sample sub-divided by
probabilistic population assignment, as described above. This Fig. 11 gives an 
indication of a systematic shift of the  mode, shifting to low metallicities 
when one goes from the thin disc stars to the thick disc stars (Fig. 11). The 
$z$-distance distribution of our sample is shown in Fig. 12. While the typical 
star is at distance $|z|\sim0.5$ kpc, there is a significant sample at
larger distances. The range in $z-$distance is large
enough to allow us to consider vertical metallicity gradient estimation.

\begin{figure}
\begin{center}
\includegraphics[scale=0.60, angle=0]{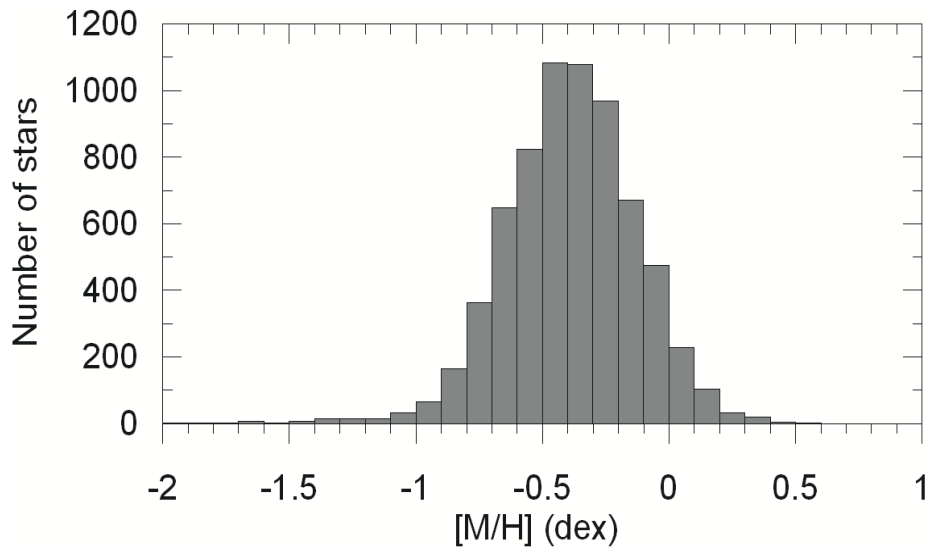}
\caption[] {The metallicity distribution for our cleaned sample of RC
 stars, using  RAVE DR3 abundances.}
\end{center}
\end{figure}

\begin{figure}
\begin{center}
\includegraphics[scale=0.60, angle=0]{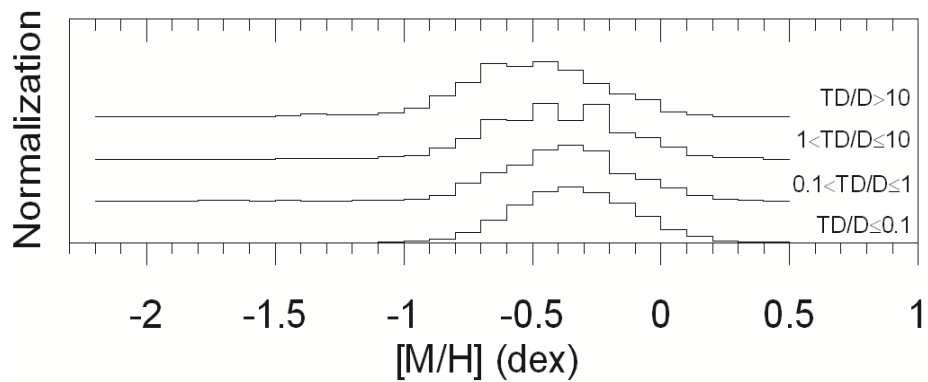}
\caption[] {The normalized metallicity distribution as a function of
  assigned (probabilistic) population types.}
\end{center}
\end{figure}

\begin{figure*}
\begin{center}
\includegraphics[scale=0.60, angle=0]{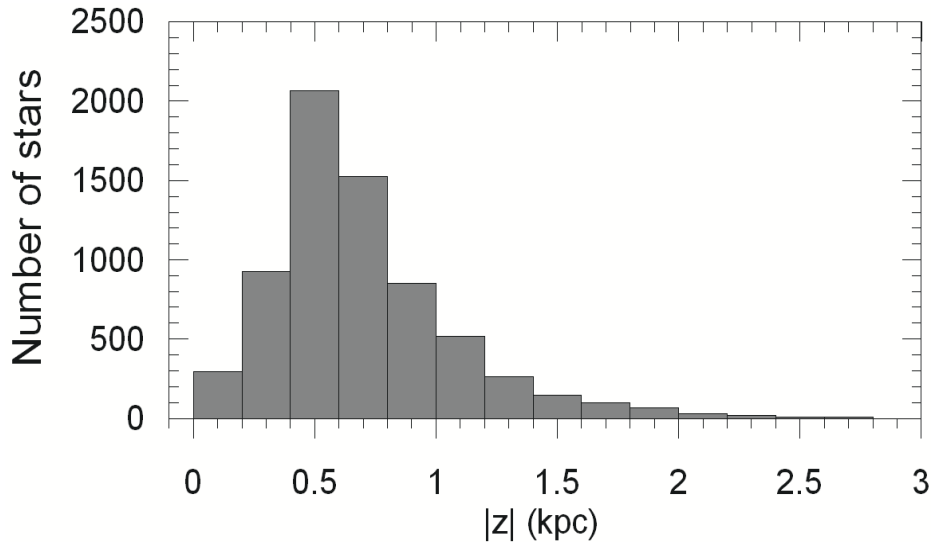}
\caption[]{Distribution of the distances from the Galactic plane for
  our red clump sample}
\end{center}
\end{figure*}

The whole sample of stars, with no consideration of population
assignment, were separated into five bins in distance, and mean
$z$-distances from the Galactic plane, mean metallicities, and mean
vertical eccentricities were calculated for each bin. These are
presented in Table 3. Fig. 13 summarises the dependence of the
results on vertical distance from the Galactic plane. The apparent
variation of the mean metallicity with $z$-distance from the Galactic
plane in Fig. 13 indicates the existence of a vertical metallicity
gradient for RC stars. This figure does not however allow
discrimination between a true gradient in a consistently defined
population (e.g. ``thick disc'') and a changing relative contribution
from two populations (e.g. ``thin disc'' and ``thick disc''), which
have different modal abundances, and for which either one, both, or
neither has an intrinsic gradient. Also given in Fig. 13 is the
variation of the vertical eccentricity with $z$, which again shows
either a smooth transition from thin disc to thick disc
eccentricities, or a changing population mix, or both disc.

\begin{table}
\center 
\caption{The mean metallicity and vertical eccentricities as a function of 
distance from the Galactic plane.}
\begin{tabular}{ccccc}
\hline
$z$ ranges & N    & $<z>$ &  $<M/H>$ & $<e_v>$ \\
(kpc)      &      & (kpc) &   (dex)  &       \\
\hline
(0.0, 0.5] & 2261 & 0.35  &  -0.33$\pm$0.25 & 0.087$\pm$0.106 \\   
(0.5, 1.0] & 3376 & 0.70  &  -0.41$\pm$0.25 & 0.161$\pm$0.136 \\
(1.0, 1.5] & 857  & 1.19  &  -0.47$\pm$0.25 & 0.271$\pm$0.187 \\
(1.5, 2.0] & 222  & 1.71  &  -0.56$\pm$0.25 & 0.378$\pm$0.222 \\
(2.0, 2.5] & 54   & 2.20  &  -0.56$\pm$0.24 & 0.460$\pm$0.228 \\
\hline
\end{tabular}  
\end{table}

\begin{figure*}
\begin{center}
\includegraphics[scale=0.50, angle=0]{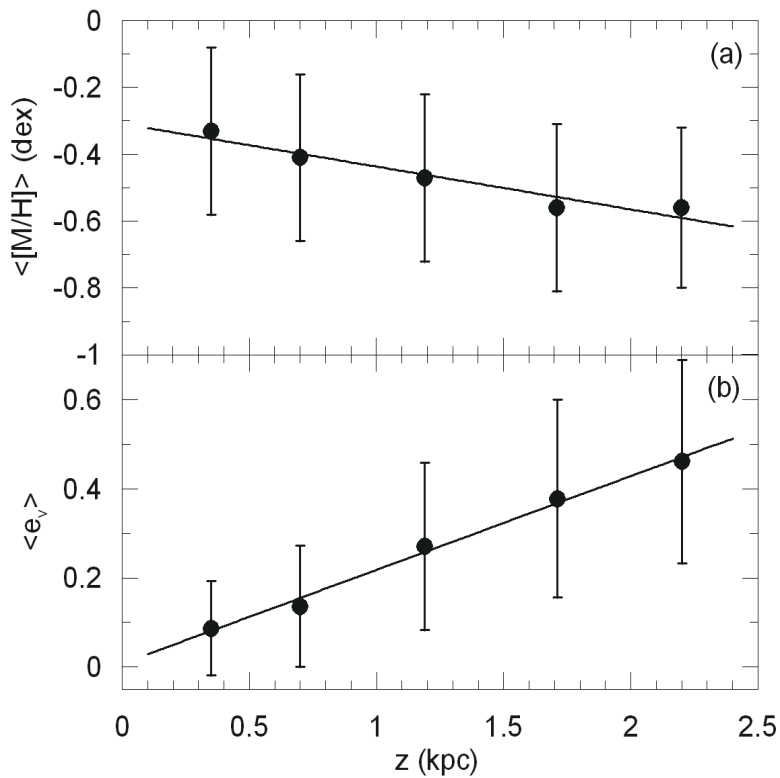}
\caption[] {Variation of the mean metallicity (top) and vertical orbital 
eccentricity (bottom) as a function of distance from the Galactic plane.}
\end{center}
\end{figure*}

\subsection{Metallicity gradients using the kinematical population assignment 
  method}
We consider the metallicities as a function of the mean orbital Galactocentric 
radial distance ($R_{m}$) and maximum distance to the Galactic plane 
($|z_{max}|$) for each different population defined in Section 4, and test for 
the presence of vertical and radial metallicity gradients for each population. 
We fitted the distributions to linear equations, whose gradient is any 
metallicity gradient, $d[M/H]/dR_{m}$ or $d[M/H]/dz_{max}$. The results are 
shown in Fig. 14. 

The radial $d[M/H]/ dR_{m}$ gradients are small or consistent with
zero. The best determined value (largest ratio of gradient value to error 
value) is for low probability thin disc stars, where the gradient is 
$d[M/H]/dR_{m}=-0.031\pm0.003$ dex kpc$^{-1}$ (Table 4). The only metallicity 
gradient consistent with zero (with small formal errors) is for low 
probability thick disc stars.

The vertical metallicity gradients are (absolutely) much larger than the
radial ones, and statistically are detected. The largest gradient in the 
vertical direction is steeper for high probability thin disc stars, relative to 
the other populations, viz $d[M/H]/dz_{max}=-0.109\pm0.008$ dex kpc$^{-1}$.
Additionally, the metallicity gradient for high probability thick disc
stars is not zero, i.e. $d[M/H]/dz_{max}=-0.034\pm0.003$ dex
kpc$^{-1}$.

\begin{table}
\center
\caption{Radial and vertical metallicity gradients for red clump RAVE stars evaluated from
kinematical and dynamical data. The meanings of the probabilistic
population assignments, $TD/D$ and $e_v$, are explained in the text.}
\begin{tabular}{lccc}
\hline
 Population type  & $d[M/H]/dR_m$    & $d[M/H]/dz_{max}$ & Sample size\\
                  & (dex kpc$^{-1}$) & (dex kpc$^{-1}$)  &            \\
\hline
$TD/D\leq 0.1$    & -0.041$\pm$0.003 & -0.109$\pm$0.008 & 3385\\
$0.1<TD/D\leq 1$  & -0.031$\pm$0.003 & -0.086$\pm$0.013 & 1151\\
$1<TD/D\leq 10$   & -0.001$\pm$0.005 & -0.036$\pm$0.016 & ~646\\
$TD/D>10$         &  0.017$\pm$0.008 & -0.034$\pm$0.003 & 1599\\
$e_v\leq0.07$     & -0.041$\pm$0.007 & -0.260$\pm$0.031 & 1269\\
$e_v\leq0.12$     & -0.025$\pm$0.040 & -0.167$\pm$0.011 & 3448\\
$0.12<e_v\leq0.25$& -0.013$\pm$0.004 & -0.103$\pm$0.008 & 2389\\
$e_v>0.25$        &  0.022$\pm$0.006 & -0.022$\pm$0.005 & ~944\\
\hline
\end{tabular}
\end{table} 

\begin{figure*}
\begin{center}
\includegraphics[scale=0.70, angle=0]{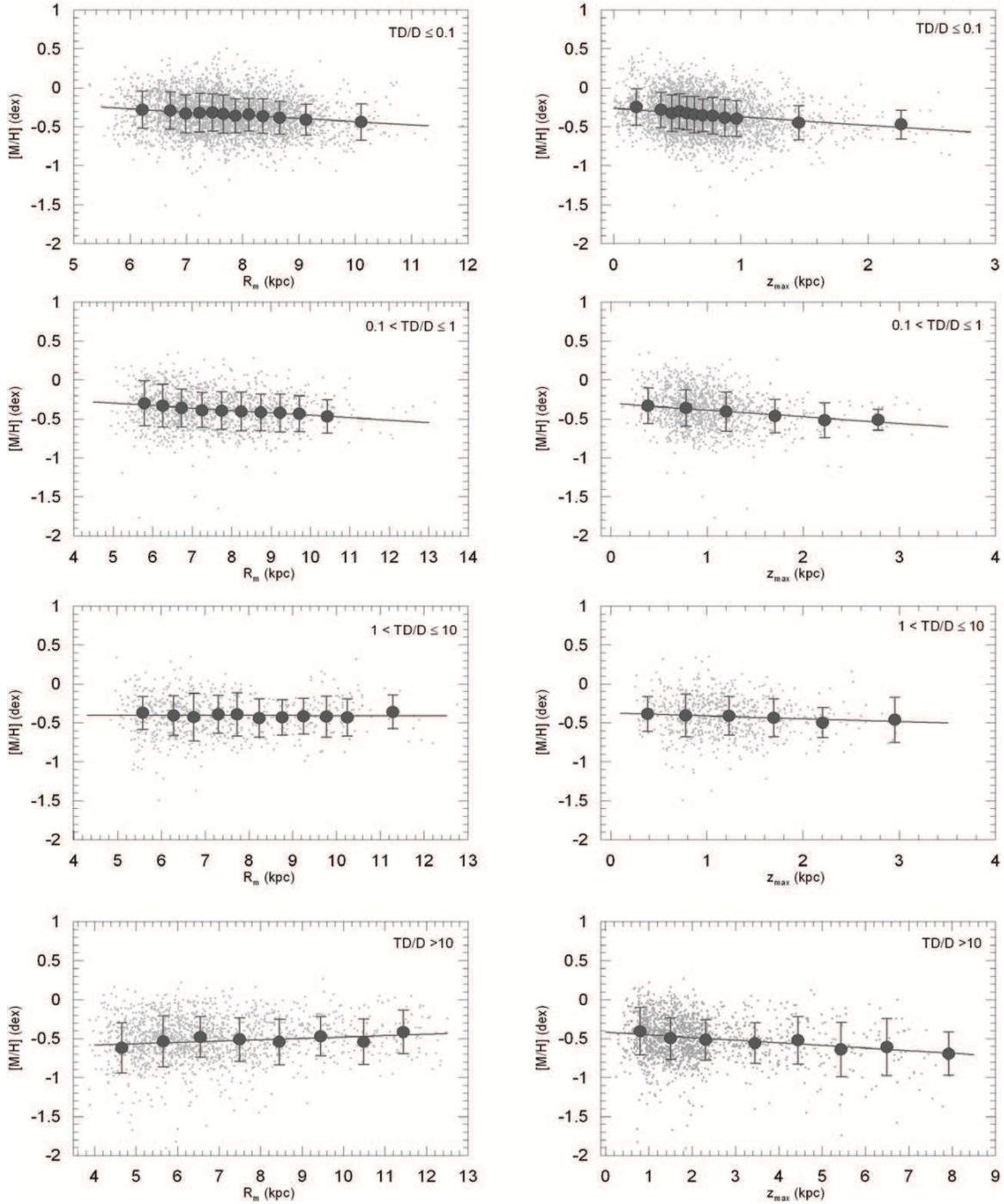}
\caption[] {Radial ($R_m$-$[M/H]$) and vertical ($z_{max}$-$[M/H]$)
  metallicity gradients for the red clump sub-samples, subdivided
into probabilistic population types as described in the text.}
\end{center}
\end{figure*}

\begin{figure*}
\begin{center}
\includegraphics[scale=0.70, angle=0]{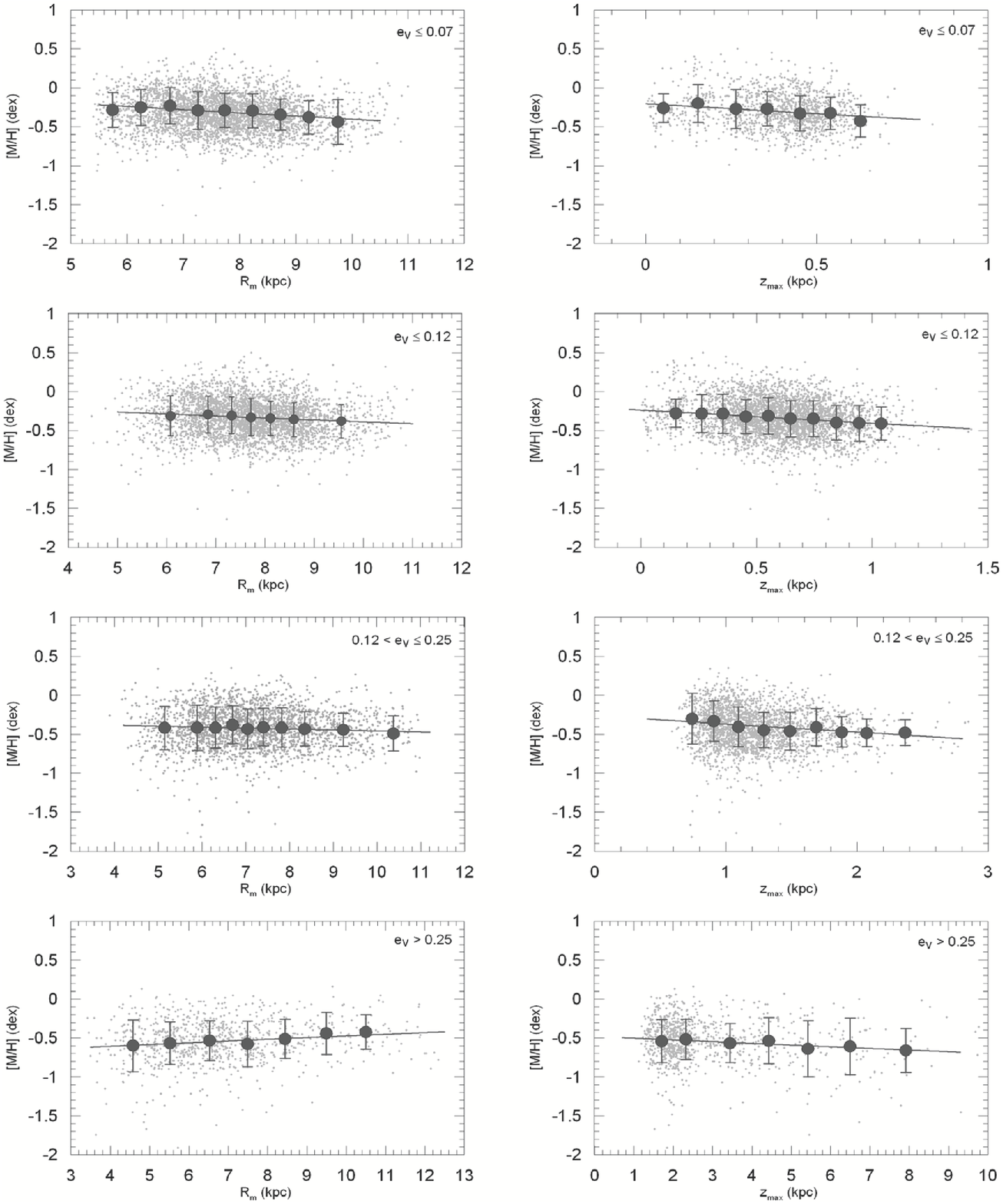}
\caption[] {Radial ($R_m$-$[M/H]$) and vertical ($z_{max}$-$[M/H]$) metallicity 
gradients for the red clump sub-samples, subdivided by vertical orbital 
eccentricity.}
\end{center}
\end{figure*}

\subsection{Metallicity gradients using the dynamical population assignment method}
We now consider the metallicities as a function of the mean orbital 
Galactocentric radial distance ($R_{m}$) and maximum distance to the Galactic 
plane ($|z_{max}|$) for the populations defined by their eccentricities in 
Section 4.2, with one slight modification, in that we add a population
defined by $e_{v}\leq0.07$ (1269 stars) representing the blue stars. We make 
this modification due to our experience in analysing the RAVE dwarf stars. In 
\citet{Coskunoglu11a}, we showed that the blue stars which have the smallest 
orbital eccentricities (most circular orbits) have also steeper metallicity 
gradients relative to samples with larger orbital eccentricities. 

The results are presented in Table~4 and Fig. 15. The steepest metallicity gradient 
is for $e_{v}\leq0.07$ in the vertical direction, i.e. 
$d[M/H]/dz_{max}=-0.260\pm0.031$ dex kpc$^{-1}$. The vertical metallicity gradient 
systematically decreases with increasing $e_{v}$ values, becoming close to zero for 
the largest vertical eccentricities,  $e_{v}>0.25$. As noted above, our (red clump 
giants) sample consists of thin and thick disc stars. The largest orbital 
eccentricities correspond to  thick disc stars. 

That is, the vertical metallicity gradient for the thick disc is 
close to zero. The trend of the radial metallicity gradient is almost the 
same as the vertical metallicity gradient; less steep. For example, 
$d[M/H]/dR_{m}=-0.041\pm0.007$ dex kpc$^{-1}$ for $e_{v}\leq0.07$. 

\section{Discussion and Conclusion}
We have used the RAVE DR3 data release to identify RC stars, further excluding 
cool stars, and those with the most uncertain space motions. We used the 
calibrated RAVE DR3 metallicities and the mean radial and maximum distances 
to investigate the presence of radial and vertical metallicity gradients, 
dividing the sample into a variety of subsamples. 

We derive significant radial and vertical metallicity gradients for high 
probability thin disc stars and for the subsample with $e_{v}\leq0.07$. We 
derive significant and marginally shallower gradients for the other 
subsamples. We do not detect any significant gradients for thick disc 
stars. Vertical metallicity gradients are much steeper than the radial ones, 
for the same subsample, as (perhaps) expected. We derive the metallicity 
gradients for the subsample $e_{v}\leq0.07$ which are the youngest sample stars, 
to be $d[M/H]/dR_{m}=-0.041\pm0.007$ dex kpc$^{-1}$ 
and $d[M/H]/dz_{max}=-0.260\pm0.031$ dex kpc$^{-1}$. 

The radial metallicity gradients for all subsamples are rather close to 
the corresponding ones obtained for F-type dwarfs. Hence, the discussion 
in \citet{Coskunoglu11a} is also valid here. We cannot 
determine any vertical metallicity gradient for RAVE dwarfs due to their 
small distances from the Galactic plane. However, we can compare our 
results with those obtained for giants. \citet{Chen11} investigated 
the metallicity gradient of the thick disc by using Red Horizontal Branch 
(RHB) stars from SDSS DR8 \citep{Aihara11} and they found two different vertical 
metallicity gradients estimated in two ways. One is a fit to the Gaussian 
peaks of the metallicity histograms of the thick disc by subtracting minor 
contributions from the thin disc and the inner halo based on the Besan\c con 
Galaxy model \citep{Robin96}. The resulting gradient is 
$d[M/H]/dz=-0.12\pm0.01$ dex kpc$^{-1}$ for $0.5<|z|<3$ kpc. The other 
method is fitting the data linearly for the stars $1<|z|<3$ kpc where
the thick disc is the dominant population. Five subgroups were then selected in 
different directions in the $X-|z|$ plane to investigate the difference in 
the vertical metallicity gradient between the Galactocentric and anti-Galactocentric
directions. They found that a vertical gradient of 
$d[M/H]/dz=-0.22\pm0.07$ dex kpc$^{-1}$ was detected for five directions except 
for one involving the population of stars from the bulge. 

Neither of the vertical metallicity gradients claimed by \citet{Chen11} are in 
agreement with our results, nor are they consistent with studies appearing in 
previous works in the literature. 

Another recent investigation of the vertical metallicity 
gradient for the thick disc is that of \citet{Ruchti11}. In that paper 
the authors gave a sample of 214 Red Giant Branch stars, 31 red
clump/red horizontal branch stars, and 74 main-sequence/sub-giant 
branch metal-poor stars. They found that the thick disc $[\alpha/Fe]$ ratios 
are enhanced, and have little variation ($<0.1$ dex). Their sample further 
allowed, for the first time, investigation of the gradients in the metal-poor thick 
disc. For stars with $[Fe/H]<-1.2$ dex, the thick disc shows very small 
gradients, $<0.03\pm0.02$ dex kpc$^{-1}$, in $\alpha$ enhancement, while 
they found a $d[Fe/H]/dR=+0.01\pm0.04$ dex kpc$^{-1}$ radial gradient 
and a $d[Fe/H]/dz=-0.09\pm0.05$ dex kpc$^{-1}$ vertical gradient in iron 
abundance. We consider only the gradient in iron abundance, not the gradient 
in $\alpha$ enhancement. We may transform  published iron abundances to RAVE 
metallicity values by means of the equation \citep{Zwitter08},
\begin{equation}
[M/H] = [Fe/H] + 0.11 [1\pm(1-e^{-3.6|[Fe/H]+0.55|})].
\end{equation}
This reveals that the \citet{Ruchti11} vertical gradient is consistent with
the vertical metallicity gradient determined here, within the errors,
for high probability thick disc stars,
$d[Fe/H]/dz_{max}=-0.034\pm0.003$ dex kpc$^{-1}$. However there is a
difference between the corresponding radial metallicity gradients in
the two studies. An explanation for this disagreement may be the
differing metallicity range of the sample stars used in the two works. 
In the present study we consider stars with $[M/H]>-1.1$ dex, with a 
minority at the metal-poor tail in our work, while the corresponding 
selection is $[Fe/H]<-1.2$ dex in \citet{Ruchti11}.
                
The radial metallicity gradients we have estimated from red
clumps giants stars are consistent with those derived in paper II for
dwarfs \citep{Coskunoglu11a}. The robust metallicity gradients we
determine are $d[M/H]/dR_{m}=-0.041\pm0.003$ dex kpc$^{-1}$ for the
high probability thin disc stars, the population type labelled
with $TD/D\leq0.1$, and $d[M/H]/dR_{m}=-0.041\pm0.07$ dex kpc$^{-1}$
for the sample with eccentricity $e_{v}\leq0.07$. Samples biased to
low probability thin disc and thick disc stars show systematically
shallower gradients. Complementary to the dwarf sample, the distance
range of the red clump giant stars, with median distance of 1.4 kpc,
provides information on vertical metallicity gradients. The vertical
metallicity gradients for the high probability thin disc stars and for
the sample with $e_{v}\leq0.07$ are $d[M/H]/dz_{max}=-0.109\pm0.008$
dex kpc$^{-1}$ and $d[M/H]/dz_{max}=-0.260\pm0.031$ dex kpc$^{-1}$,
respectively. For high probability thick disc stars we could detect a
vertical metallicity gradient; a shallow one however,
i.e. $d[M/H]/dz_{max}=-0.034\pm0.003$ dex kpc$^{-1}$.

From our analysis, we may conclude that, despite their greater distances 
from the Galactic plane, the RAVE DR3 red clump giant stars confirm the radial 
metallicity gradients found for RAVE DR3 dwarf stars. Because of their greater 
distances from the Galactic plane, the RAVE DR3 RC stars also permit 
vertical metallicity gradients to be measured. These findings can be used to 
constrain formation scenarios of the thick and thin discs.

\section{Acknowledgments}
We would like to thank the referee Dr. Martin L\'opez-Corredoira for
his useful suggestions that improved the readability of this paper.

Funding for RAVE has been provided by: the Australian Astronomical Observatory; 
the Leibniz-Institut fuer Astrophysik Potsdam (AIP); the Australian National 
University; the Australian Research Council; the French National Research 
Agency; the German Research Foundation; the European Research Council 
(ERC-StG 240271 Galactica); the Istituto Nazionale di Astrofisica at Padova; 
The Johns Hopkins University; the National Science Foundation of the USA 
(AST-0908326); the W. M. Keck foundation; the Macquarie University; the 
Netherlands Research School for Astronomy; the Natural Sciences and 
Engineering Research Council of Canada; the Slovenian Research Agency; the 
Swiss National Science Foundation; the Science \& Technology Facilities 
Council of the UK; Opticon; Strasbourg Observatory; and the Universities of 
Groningen, Heidelberg and Sydney. The RAVE web site is at 
http://www.rave-survey.org 

This publication makes use of data products from the Two Micron All
Sky Survey, which is a joint project of the University of
Massachusetts and the Infrared Processing and Analysis
Center/California Institute of Technology, funded by the National
Aeronautics and Space Administration and the National Science
Foundation.  This research has made use of the SIMBAD, NASA's
Astrophysics Data System Bibliographic Services and the NASA/IPAC
ExtraGalactic Database (NED) which is operated by the Jet Propulsion
Laboratory, California Institute of Technology, under contract with
the National Aeronautics and Space Administration.

\end{document}